\documentclass[letter]{aa}  

\usepackage[colorlinks=true,linkcolor=blue,citecolor=blue,filecolor=blue,urlcolor=blue]{hyperref}
\usepackage{graphicx}
\usepackage{txfonts}

\begin{document}

   \title{The flaring drill in the Galactic centre}

   \subtitle{Did the IRS 13 cluster carve out the mini-cavity in the mini-spiral?}

   \author{Jaroslav Haas\inst{1}\thanks{\email{haas@sirrah.troja.mff.cuni.cz}},
           Pavel Kroupa\inst{1,2}, Florian Pei\ss{}ker\inst{3},
           \and Mark R. Morris\inst{4}}

   \institute{Charles University, Faculty of Mathematics and Physics, Astronomical Institute,
              V Hole\v{s}ovi\v{c}k\'ach 2, CZ-18000 Prague, Czech Republic
              \and
              Helmholtz-Institut f\"{u}r Strahlen- und Kernphysik, University of Bonn,
              Nussallee 14-16, D-53115 Bonn, Germany
              \and
              I. Physikalisches Institut der Universit\"{a}t zu K\"{o}ln, Z\"{u}lpicher Str. 77,
              D-50937 K\"{o}ln, Germany
              \and
              University of California, Department of Physics and Astronomy, Los Angeles,
              CA 90095-1547, USA
             }

   \abstract
   {The mini-cavity is a low-density region observed in the complex of streams
   of ionized gas around the Galactic central supermassive black hole, Sgr~A$^\star$, known as
   the mini-spiral. Its near-circular shape is suggestive of a formation due to the effect
   of stellar winds. No suitable stars are currently observed within the mini-cavity, however.}
   {In this study we assessed whether the mini-cavity could have been formed by the winds of the
   stars from the neighbouring IRS~13 cluster that were located at the position of the mini-cavity
   in the past but moved away from it later on owing to their orbital motions around
   Sgr~A$^\star$. Furthermore, we estimated the rate of accretion of the then-abundant
   interstellar medium onto the putative intermediate-mass black hole that has been proposed to
   reside in the IRS~13 cluster and the corresponding X-ray luminosity of this black hole.}
   {The estimates were obtained analytically using the astrophysical properties reported
   for the involved objects and the environment.}
   {Based on our results, we suggest that the mini-cavity was formed by the winds of the IRS~13
   cluster member stars about 300 years ago, when this cluster went through the Bar region of the
   mini-spiral. The accompanying accretion of the interstellar medium onto the
   putative intermediate-mass black hole in this cluster may have produced multiple X-ray flares
   with luminosities of $\approx10^{39}~\mathrm{erg~s}^{-1}$. Such flares are compatible with the
   X-ray reflections currently observed on the molecular clouds in the complexes Sgr~A, B, and C,
   including the necessary light-travel time delay.}
   {}
   \maketitle
\section{Introduction}
\label{intro}
Observations of the ionized gas in the Sgr~A West sector of the Galactic centre revealed
a set of streams widely referred to as the mini-spiral
\citep{Ekers83,Lo83,Zhao09,Zhao10,Nitschai20}. It comprises the Northern, Eastern, and Western
Arms and the Bar, which appears to be a region where the Northern and Eastern Arms intersect
\citep{Liszt03,Zhao09}. The particle density within the mini-spiral  ($n_\mathrm{ms}$)
is reported to be $\approx3\times10^4~\mathrm{cm}^{-3}$
\citep{Haggard24}, which corresponds to a mass density
($\rho_\mathrm{ms}$) of $\approx740\,M_\odot~\mathrm{pc}^{-3}$ if purely hydrogen gas is assumed.
\begin{figure}
\centering
\includegraphics[width=\columnwidth]{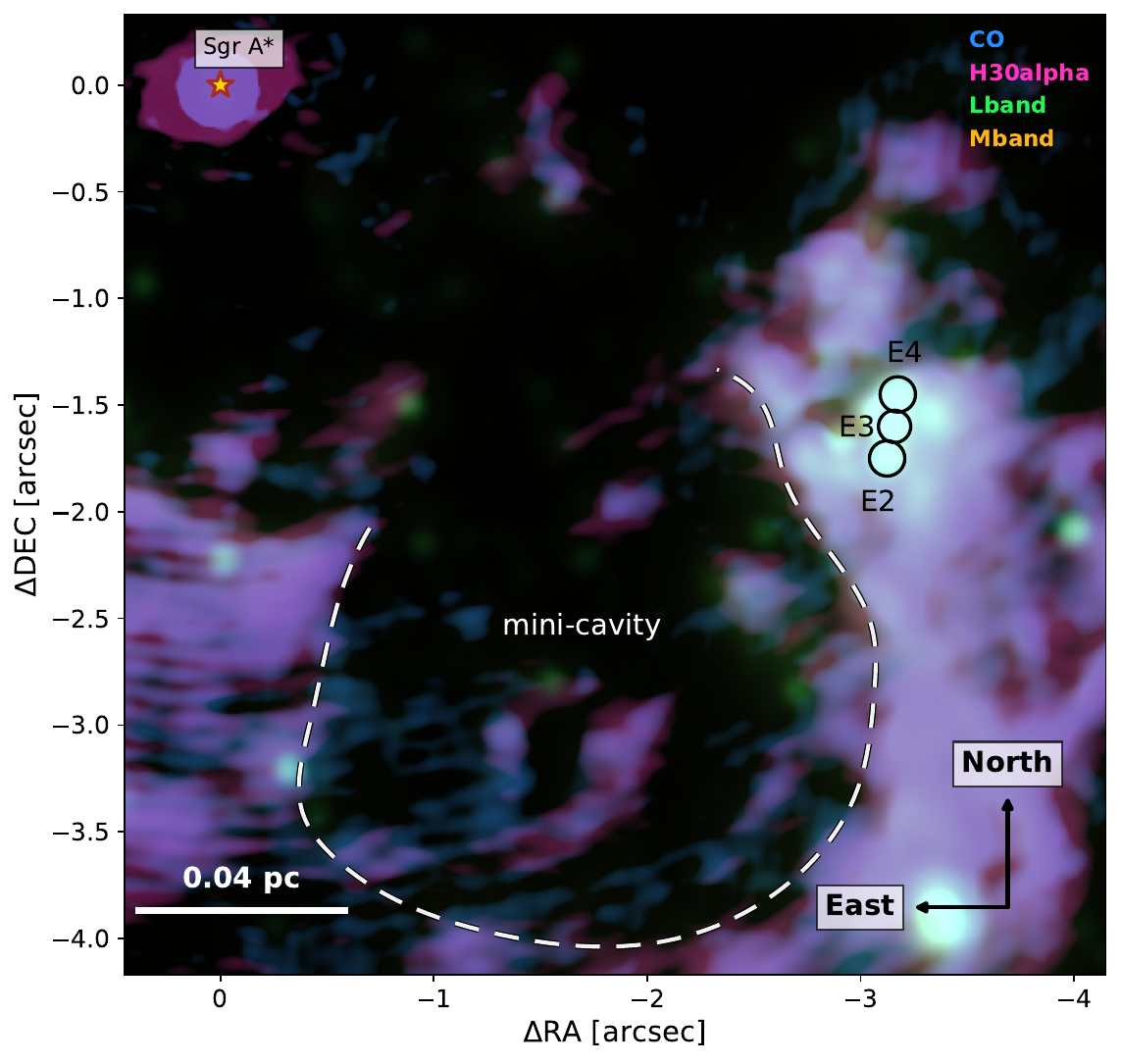}
\caption{Polychromatic finding chart for the subject region. The axes show offsets from
Sgr~A$^\star$ (top-left corner) in right ascension, $\Delta\mathrm{RA}$, and declination,
$\Delta\mathrm{DEC}$. Stars E2 and E4, commonly associated with the IRS~13 cluster, and the
putative IMBH source E3 are indicated by the black circles. The mini-cavity is denoted by
the dashed curve. The data used here include wavelengths 3.6--$4.5~\mu\mathrm{m}$ and
observations of $^{12}$CO at 343~GHz and H30$\alpha$ at 232~GHz.
For a detailed description of the data, see \citet{Peissker23a}.}
\vspace*{-2mm}
\label{chart}
\end{figure}

In the Bar region, 
there is a near-circular, low-density region with a radius of $\approx0.04$~pc commonly called the
mini-cavity \citep[see Fig.~\ref{chart} here and][]{Morris87,YusefZadeh89,Eckart92,Zhao09}; it is located at a projected distance of $\approx0.14$~pc
south-west of the central supermassive black hole Sgr~A$^\star$ \citep{GRAVITY22}. Various formation mechanisms for the mini-cavity have been suggested
(see \citealt{Zhao09} and \citealt{Ferriere12} for reviews). Because of its near-circular shape, an obvious choice
would be the effect of stellar wind blown by a star located within the mini-cavity or a
supernova explosion. The latter is not likely, owing to the lack of the characteristic
non-thermal radiation emanating from the mini-cavity \citep{Eckart92}. The former is somewhat
complicated because no wind-blowing star is currently observed within the mini-cavity. Hence,
the stellar winds from the stars in its surroundings, either direct
\citep{Morris87} or collimated by the gravity of Sgr~A$^\star$ \citep{Wardle92,Melia96}, or the
wind from the accreting Sgr~A$^\star$ itself \citep{Eckart92} have been invoked.

However, these studies did not take the kinematics of the stars near the
mini-cavity into account, and the wind-blowing stars may have already left the mini-cavity due to these kinematics. The most
straightforward candidates for such stars are those
observed in the so-called IRS~13 cluster \citep{Maillard04,Paumard06,Peissker23,Peissker24}, which is located just outside the north-western rim
of the mini-cavity. Two of these stars, designated as E2 and E4, are Wolf-Rayet stars with
powerful winds \citep{Martins07,Zhu20}.
In this Letter we investigate the possibility that the
mini-cavity was formed by the stellar winds of these stars in the past, during their passage
through the Bar region of the mini-spiral. We begin with an assessment of the observed motions
of E2 and E4 with respect to the overall kinematics of the ambient medium in the Bar
region (Sect.~\ref{penetration}) and then evaluate the impact of their winds on this medium
(Sect.~\ref{formation}).

The IRS~13 cluster has been proposed to host an intermediate-mass black hole\footnote{The
Galactic centre may host a population of IMBHs \citep{Haas26}. These IMBHs can seed the
formation of new star clusters in their accretion discs, which could explain the disc-like
arrangement of young stellar objects in the IRS~13 cluster reported by \citet{Peissker23}.}
\citep[IMBH;][]{Maillard04,Tsuboi17,Peissker24} in order to be gravitationally bound
(including E2 and E4). In Sect.~\ref{accretion} we estimate the possible
accretion rate onto such a black hole and its X-ray luminosity during the penetration of the
mini-spiral Bar region by the whole IRS~13 cluster. In Sect.~\ref{reflections}
we assess whether this luminosity can explain the X-ray flare reflections observed on the
molecular clouds in the Sgr~A, B, and C sectors of the Galactic centre
\citep{Clavel13,Ponti10,Terrier10,Ponti14,Chuard18,Terrier18,Marin23}. We discuss our
results in Sect.~\ref{discussion}.
\vspace*{-2mm}
\section{The mini-spiral penetration by the IRS~13 cluster}
\label{penetration}
Though there is no clear consensus in the literature about the nature of the
IRS~13 cluster and its member stars, the two Wolf-Rayet stars E2 and E4 are
widely assumed to be part of it (regardless of whether this cluster is gravitationally bound
or not). The directions of the observed
motion of these stars in the plane of the sky are compatible with their location within the
mini-cavity in the past. The actual velocity of this motion of $\approx200$~km/s
\citep{Wang20,Zhu20,Peissker23} thus suggests that E2 and E4 were near the centre
of the mini-cavity ($\approx0.06$~pc in projection from their current locations) roughly 300 years
ago.

The kinematics of the mini-spiral gas around the mini-cavity is
rather complex \citep{Zhao09,Zhao10,Nitschai20}, but the gas shows a westward bulk motion in the
plane of the sky. Its direction and velocity are similar to those of E2 and E4.
The line-of-sight velocity of the gas flow in the vicinity of the mini-cavity ranges
from $\approx-100$~km/s to $\approx+100$~km/s. The currently observed line-of-sight velocity
of E2 is uncertain: $\approx-50$~km/s \citep{Zhu20} or $\approx+60$~km/s \citep{vonFellenberg22}.
For E4, $\approx+70$~km/s is reported \citep{Zhu20,vonFellenberg22}.

Based on this simple analysis, we propose that about 300 years ago, the two Wolf-Rayet stars E2
and E4 went through the mini-spiral at the location of the mini-cavity. We assume that the
relative velocity of
the stars with respect to the gas, which was then abundant in this region, was $\approx100$~km/s.

Furthermore, if the IRS~13 cluster were bound by an IMBH
\citep{Maillard04,Tsuboi17,Peissker24}, the observed motions of E2 and E4 would
include their orbits around this IMBH. The impact of such orbital motions on the timing and
plausibility of the penetration event would depend not only on the mass of the IMBH and the
particular parameters of the E2 and E4 orbits around it, but also on the orbit of the IRS~13 cluster
as a whole around Sgr~A$^\star$. None of these are currently well constrained, to our
knowledge\footnote{See, for example, the conflicting reported motions of source E3, which is
commonly associated with the putative IMBH \citep{Schoedel09,Zhu20,Tsuboi22,Peissker23}.}. Hence,
for simplicity, we assumed in this study that the timing of the penetration event is not
affected by the presence of the putative IMBH and that its relative velocity with respect to
the mini-spiral gas during the penetration event was also the above proposed $\approx100$~km/s.
\vspace*{-2mm}
\section{The formation of the mini-cavity by stellar winds}
\label{formation}
To assess the impact of the winds of the two Wolf-Rayet stars E2 and E4 on the
surrounding medium during their passage through the mini-spiral in the Bar region, we
adopted the approach of \citet[see our Appendix~\ref{bubble-growth}]{Weaver77}.
According to \citet{Zhu20}, the mass loss rate of both E2 and E4 can be estimated as
$\mathrm{d}m_\mathrm{star,E2,E4}/\mathrm{d}t\approx1.8\times10^{-5}\,M_\odot~\mathrm{yr}^{-1}$.
The wind terminal velocities for these two
stars differ, $v_\mathrm{term,E2}\approx750$~km/s versus $v_\mathrm{term,E4}\approx2200$~km/s.
If we assume that the mini-cavity was blown to its currently observed dimensions
($\approx0.04$~pc in radius) by the stronger wind of E4, which was located at its centre in the
past, the necessary time as given by Eq.~(\ref{bubble}) is $\approx120$ years for the mini-spiral
gas density $\rho_\mathrm{ms}\approx740\,M_\odot~\mathrm{pc}^{-3}$ (see Sect.~\ref{intro}).
This time serves as an estimate of the duration of the penetration event and holds
even if we consider that E4 and E2 act together because the wind
mechanical luminosity (Eq. \ref{mechanical}) for E2 is about a factor of 9 lower than for E4.
Such a duration estimate is, however, an upper limit as the mini-cavity
would have continued to expand even after the stars had left it.

Considering the relative velocity of the stars with respect to the gas of $\approx100$~km/s
(estimated in Sect.~\ref{penetration}), the stars covered a distance of $\approx0.01$~pc within
the mini-spiral Bar region during the penetration event of the estimated duration.
This distance represents an estimate of the local line-of-sight
thickness of the gas stream, which is thus smaller than its width in the plane
of the sky, $\approx0.1$~pc \citep[see Fig.~1 in][]{Zhao09}.

\begin{figure}
\centering
\includegraphics[width=0.8\columnwidth]{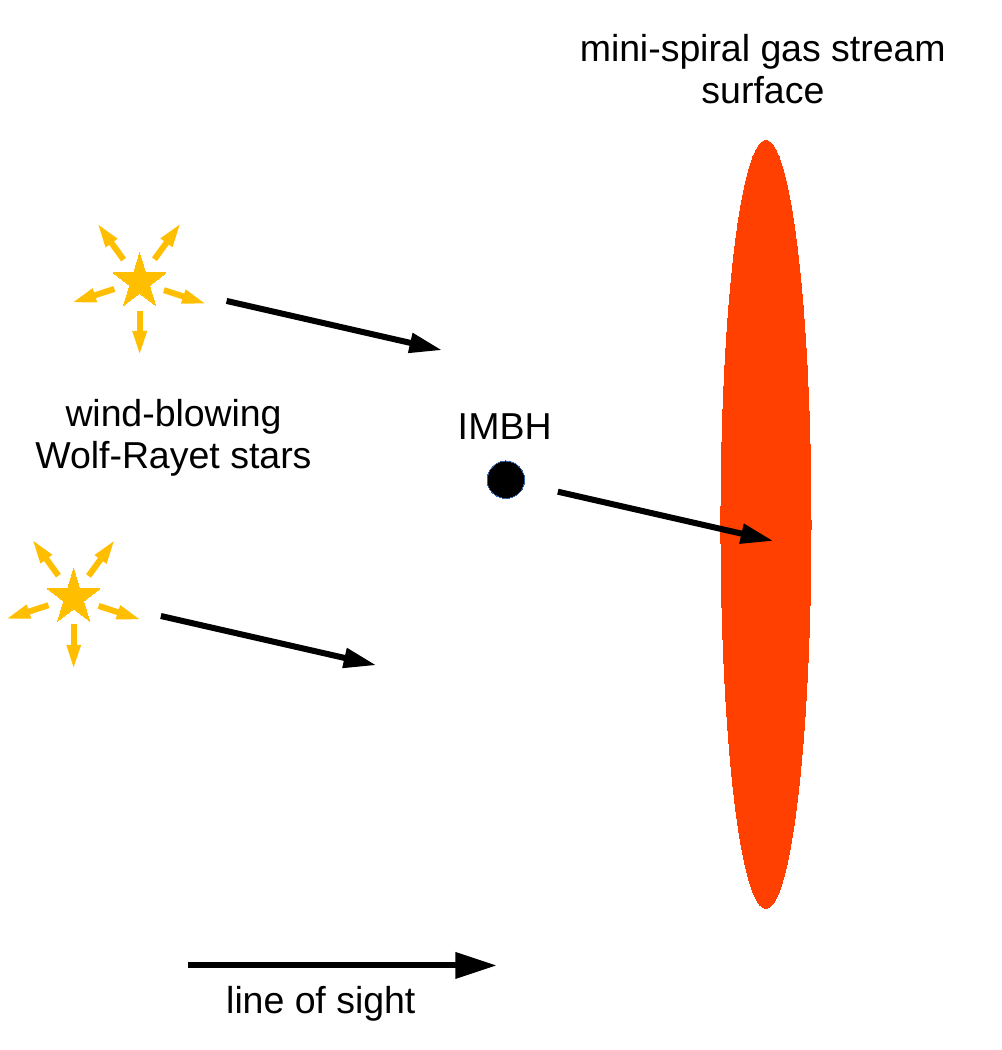}
\caption{Most favourable geometry for the model presented here. The
mini-spiral gas stream is rather thin and penetrated roughly along the line of sight. The
IMBH is ahead of the wind-blowing Wolf-Rayet stars.}
\vspace*{-2mm}
\label{geometry}
\end{figure}

The average expansion velocity of the bubble, calculated from the observed radius
of the mini-cavity and the above penetration event duration, is $\approx330$~km/s.
This is compatible with the mini-cavity being near-circular as such a velocity significantly
exceeds the assumed relative velocity of the stars and the gas ($\approx100$~km/s)
during the event. For relative velocities higher than the bubble
expansion velocity, an elongated cavity would be expected, except for the case of
penetrations roughly along the line of sight. Such high-angle penetrations (see
Fig.~\ref{geometry}) are thus the most favourable for the model presented in this Letter. They
are also consistent with the similar westward motion in the plane of the sky observed for
the stars and for the mini-spiral gas in the Bar region (see Sect.~\ref{penetration}).

The formation of a near-circular cavity is also more likely for penetrations of
a rather thin gas layer with a thickness smaller than the radius of the stellar wind-blown
bubble. This is in accord with the observed radius of the mini-cavity of $\approx0.04$~pc
being about a factor of 4 larger than the gas stream thickness estimate obtained in this section.
\vspace*{-2mm}
\section{Accretion onto the putative intermediate-mass black hole in the IRS~13 cluster}
\label{accretion}
As described in Sect.~\ref{formation}, the stellar winds of the
two Wolf-Rayet stars from the IRS~13 cluster created a low-density bubble in the ambient medium
during their penetration of the mini-spiral. Hence, if the putative IMBH in this cluster were
close to these stars, it would soon have had essentially nothing to accrete after the bubble
front shock reached it.

We thus investigated the accretion of the still
undisturbed mini-spiral medium onto the IMBH assuming a favourable geometry in which
the IMBH is located ahead of the two Wolf-Rayet stars in the direction of the mini-spiral
penetration (see Fig~\ref{geometry}). A significant accretion rate is ensured if the
distance of the IMBH from the Wolf-Rayet stars is about the same as the
mini-cavity radius of $\approx0.04$~pc, meaning the IMBH would have stayed outside of the
bubble during the entire penetration event. For the accretion,
we adopted the Bondi-Hoyle-Lyttleton accretion model (see Appendix~\ref{rate-luminosity}).

According to \citet{Peissker24}, the IMBH in IRS~13 could be as massive as
$3\times10^4\,M_\odot$. \citet{Zhu20} conclude, however, that if there is an IMBH in
IRS~13, its mass, $m_\mathrm{IMBH}$, does
not exceed a few $10^3\,M_\odot$. Here, we assumed $m_\mathrm{IMBH}=10^3\,M_\odot$ to obtain
a conservative estimate of the accretion rate.

We set the relative velocity of the IMBH with respect
to the mini-spiral medium, $v_\mathrm{IMBH-ms}$,  during the penetration event to $\approx100$~km/s based on the estimates given in Sect.~\ref{penetration}.
With such a fast relative motion, we can assume that the thermodynamics of the mini-spiral
is not important for the accretion, $v_\mathrm{IMBH-ms}\gg{}c_\mathrm{s}$.

The accretion rate (Eq. \ref{bondi}) onto the IMBH for the values of $m=m_\mathrm{IMBH}$,
$\rho=\rho_\mathrm{ms}$, and $v=v_\mathrm{IMBH-ms}$ gives
$\mathrm{d}m_\mathrm{IMBH}/\mathrm{d}t\approx2\times10^{-7}\,M_\odot~\mathrm{yr}^{-1}$.
The resulting bolometric accretion luminosity (Eq. \ref{luminosity}) reaches
$L_\mathrm{IMBH}\approx10^{39}~\mathrm{erg~s}^{-1}$ if $\epsilon=0.1$ is assumed.
If we considered the more massive IMBH according to \citet{Peissker24} with
$m_\mathrm{IMBH}\approx3\times10^4\,M_\odot$, the corresponding bolometric accretion
luminosity would be $L_\mathrm{IMBH}\approx10^{42}~\mathrm{erg~s}^{-1}$.

The spectral energy distribution depends on the details of the accretion process, but we can
roughly assume that 5--10\% of the bolometric luminosity is radiated in X-rays
\citep[see Fig. 12 in][]{Peissker24}. Hence, the resulting X-ray luminosity,
$L_\mathrm{X,IMBH}$, of the putative IMBH in the IRS~13 cluster during the penetration event
can be estimated as
$L_\mathrm{X,IMBH}\approx5\times10^{37}$--\,$10^{41}~\mathrm{erg~s}^{-1}$ for
$m_\mathrm{IMBH}\approx10^3$--\,$3\times10^4\,M_\odot$.
\vspace*{-2mm}
\section{X-ray reflections on the molecular clouds in Sgr~A, B, and C}
\label{reflections}
The interval obtained  above for the X-ray luminosity of the putative IMBH in the IRS~13
cluster during its penetration of the mini-spiral includes the value of about
$10^{39}~\mathrm{erg~s}^{-1}$. This value
was derived for the two accretion flares of Sgr~A$^\star$ suggested to explain the observed
X-ray reflections on the molecular clouds in Sgr~A, B, and C \citep{Clavel13,Chuard18,Marin23}.
The supermassive black hole Sgr~A$^\star$ was proposed
as the natural first-choice candidate for the accreting object, however without any specific
explanation for the approach of the accreted matter. While this certainly does not make
Sgr~A$^\star$ less likely due to the overall dynamics of the stars and gas in its close vicinity,
other candidates for the accreting object that fulfil the observational constraints merit
consideration.

The separation of Sgr~A$^\star$ and the mini-cavity --- and thus also the IRS~13 cluster
during its mini-spiral penetration --- is negligibly small relative to the distances of the
molecular clouds in Sgr~A, B, and C from Sgr~A$^\star$. The overall geometry of the X-ray
reflections is thus the same for the accretion onto the putative IMBH in the IRS~13 cluster.
Furthermore, the beginning of the
mini-spiral penetration estimated to have occurred
$\approx300$~years ago (see Sect.~\ref{penetration}) is in fair agreement with the dating of the accretion event responsible for
the longer of the two observed reflected flares to $\approx240$~years ago \citep{Chuard18}.

The duration of this longer reflected flare was estimated to be about a decade
\citep{Chuard18}, which is shorter than the duration of the mini-spiral penetration
by the IRS~13 cluster derived in Sect.~\ref{penetration}. The second (shorter) reflected
flare was caused by an accretion event that occurred about a century after the first one,
that is, at about the time of the end of the mini-spiral penetration. Multiple
short flares during the mini-spiral penetration, rather than one long flare that
lasted for its entire duration, are expected due to the varying accretion
flow onto the IMBH caused by the likely inhomogeneities in the matter distribution
in the mini-spiral streams.
\vspace*{-3mm}
\section{Discussion}
\label{discussion}
We have thus far formulated a formation scenario for the observed mini-cavity
that relies on the effect of the stellar winds of the Wolf-Rayet stars from the IRS~13 cluster
during their past penetration of the Bar region of the mini-spiral.
While this idea is qualitatively straightforward, the particular numerical estimates
are based on a simplified model of a spherically symmetric wind blown by a star
at rest with respect to the ambient medium. This assumption is formally not fulfilled, because of
the motion of the IRS~13 stars relative to the medium with a velocity of $\approx100$~km/s
(Sect.~\ref{penetration}). However, the mini-cavity average expansion velocity of
$\approx330$~km/s derived within this model (Sect.~\ref{formation}) is about a factor of 3
higher, which suggests that the simplification is still acceptable for the rough estimates
presented in this Letter.

The interaction of the high-velocity stellar winds of the IRS~13 stars with the ambient
medium during the mini-spiral penetration should manifest as an observable shock wave
around the mini-cavity. No such feature, however, has been reported \citep{Zhao09,Zhao10}.
We attribute this to the fact that the penetration event started $\approx300$ years ago
(Sect.~\ref{penetration}) and its duration was only $\approx120$ years
(Sect.~\ref{formation}). Subsequently, after some period of time of a purely inertia-driven
propagation, the shock wave gradually dispersed.

The gas column density of the western wall of the mini-cavity is higher than elsewhere
around its rim (see Fig.~\ref{chart}).
This is naturally explained by the fact that the winds of the IRS~13 stars
ploughed an additional portion of the gas from the Bar towards the western wall after the
penetration event due to their westward motion from the mini-cavity to their currently observed
locations.

The accretion rate onto the putative IMBH in the IRS~13 cluster estimated in
Sect.~\ref{accretion} does not take the effect of the stellar winds into account  and so
should be considered an upper limit. A detailed modelling of the accretion process
that includes both the stellar winds (and feedback in general) and the relative motion of IRS~13
with respect to the ambient medium is required to evaluate by how much the true accretion rate
is reduced. The internal dynamics of the stars within the IRS~13 cluster is also
important as it affects the relative positions of the wind-blowing stars and the accreting
IMBH during the penetration event.

Instead of an IMBH, the IRS~13 cluster might contain a segregated sub-cluster of stellar-mass
black holes of the same total mass as that of the purported IMBH, possibly representing an example of
the so-called dark star clusters \citep{Banerjee11,Wu24,RostamiShirazi24}. In such a case, the
total accretion luminosity of the black hole sub-cluster would depend on its compactness and
mass function. In the most favourable case of a very compact black hole sub-cluster
with a total mass of $\approx3\times10^4\,M_\odot$, the resulting X-ray luminosity can still be
close to the upper bound of the interval for the IMBH derived in Sect.~\ref{accretion}.
The lower bound of this interval corresponding to a rather extended, low-density black
hole sub-cluster with a total mass of $\approx10^3\,M_\odot$, however, would decrease by
about a factor of the total number of the black holes in the sub-cluster. This is because such
a sub-cluster represents a set of independent accretors rather than a single accretor of the
aggregate mass and because the accretion rate~(Eq. \ref{bondi}) depends quadratically on the mass of
the accreting object \citep[for a more detailed discussion, see][]{Haas26}.
\vspace*{-2mm}
\section{Conclusions}
\label{conclusions}
In this Letter we have formulated a novel formation scenario for the so-called mini-cavity,
a near-circular, low-density region observed in the Bar of the mini-spiral located in the
Sgr~A West sector of the Galactic centre. We propose that the mini-cavity was formed by the
stellar winds of the stars from the neighbouring IRS~13 cluster that went through the Bar about
300 years ago.
Owing to the fact that the expansion velocity of the wind-driven shock was about a factor of 3
higher than the relative velocity of the stars with respect to the ambient medium during the
penetration event, the mini-cavity acquired its near-circular shape, which it has kept to the present day.

We further argue that the accretion of the inhomogeneously distributed medium onto the putative
intermediate-mass black hole in the IRS~13 cluster during the Bar penetration
resulted in two or more X-ray flares with luminosities of $\approx10^{39}~\mathrm{erg~s}^{-1}$.
We suggest that these flares are now observed as reflections on the nearby molecular clouds in
Sgr~A, B, and C.

Accurate measurements of the three-dimensional motions of all of the IRS~13 stars (including the
source E3 commonly associated with the putative intermediate-mass black hole) and mapping
the flow of the interstellar medium around the mini-cavity are essential to test the hypothesis
presented in this Letter. A detailed modelling of the Bar penetration and the accompanying
accretion event in its complexity is also a necessary complementary step.
\begin{acknowledgements}
We thank the anonymous referee for their useful comments that helped to improve this manuscript.
JH and PK acknowledge support from the Czech Science Foundation grant No. 26-21774S.
PK acknowledges support through the DAAD Eastern-European Bonn-Prague exchange programme.
\end{acknowledgements}

\begin{appendix}
\section{Bubble growth}
\label{bubble-growth}
The model of \citet{Weaver77} assumes that
the wind of a star at rest with respect to the ambient medium of a constant mass
density, $\rho$, is steady and spherically symmetric. The time-dependent radius,
$R\left(t\right)$, of the wind-blown bubble in this model can be written as
\begin{equation}
\label{bubble}
R\left(t\right)=\alpha\left(\frac{L_\mathrm{mech}}{\rho}\right)^{1/5}t^{3/5}\,,
\end{equation}
with $\alpha$ being a numerical constant, $\alpha\approx1$, and $L_\mathrm{mech}$ the mechanical
luminosity of the wind given by
\begin{equation}
\label{mechanical}
L_\mathrm{mech}=\frac{1}{2}\frac{\mathrm{d}m_\mathrm{star}}{\mathrm{d}t}v_\mathrm{term}^2\,,
\end{equation}
where $\mathrm{d}m_\mathrm{star}/\mathrm{d}t$ is the mass loss rate and $v_\mathrm{term}$
the wind terminal velocity for the blowing star.
\section{Accretion rate and luminosity}
\label{rate-luminosity}
In the Bondi-Hoyle-Lyttleton accretion model \citep{Hoyle39,Bondi52},
the matter accretion rate, $\mathrm{d}m/\mathrm{d}t$, onto the central object of mass $m$ is
given by
\begin{equation}
\label{bondi}
\frac{\mathrm{d}m}{\mathrm{d}t}\approx4\pi\rho
\frac{\left(Gm\right)^2}{\left(v^2+c_\mathrm{s}^2\right)^{3/2}}~,
\end{equation}
where $\rho$ is the mass density of the medium in the reservoir far away from the
accreting object, $c_\mathrm{s}$ is the speed of sound in the undisturbed medium, $v$ the
relative velocity of the object with respect to the reservoir and $G$ the gravitational
constant. Note that while the Bondi-Hoyle-Lyttleton approach certainly represents a
simplification of the real accretion process, it is suitable for assessing the accretion
onto a point-like IMBH.

The corresponding bolometric accretion luminosity, $L$, can be expressed as
\begin{equation}
\label{luminosity}
L=\epsilon\frac{\mathrm{d}m}{\mathrm{d}t}c^2\,,
\end{equation}
with $\epsilon$ being the radiation generation efficiency and $c$ the speed of light.
\end{appendix}

\end{document}